\documentclass[10pt,twocolumn,letterpaper]{article}

\usepackage{wacv}
\usepackage{times}
\usepackage{epsfig}
\usepackage{graphicx}
\usepackage{amsmath}
\usepackage{amssymb}

\usepackage{booktabs}       
\usepackage{amsfonts}       
\usepackage{nicefrac}       
\usepackage{microtype}      
\usepackage{xcolor}         
\usepackage{mathrsfs}  
\usepackage{bm}
\usepackage{amssymb,amsmath}

\usepackage{subcaption}
\usepackage{wrapfig}
\usepackage{picinpar}
\usepackage{cutwin}
\usepackage{graphicx,multicol}
\usepackage{algorithm}  
\usepackage{algorithmic}
\usepackage{stmaryrd}

\usepackage{multirow}
\usepackage{algorithm}  
\usepackage{algorithmic}
\usepackage{amsfonts}
\usepackage{amsfonts,amssymb} 
\usepackage{verbatim}
\usepackage{url}
\usepackage{enumitem}

\usepackage[pagebackref=true,breaklinks=true,letterpaper=true,colorlinks,bookmarks=false]{hyperref}

\wacvfinalcopy 

\def\wacvPaperID{1762} 

\ifwacvfinal\fi
\begin{document}

\title{BirdSoundsDenoising: Deep Visual Audio Denoising for Bird Sounds}


\author{Youshan Zhang \\
Yeshiva University, NYC, NY\\
{\tt\small youshan.zhang@yu.edu}
\and
Jialu Li \\
Cornell University, Ithaca, NY\\
{\tt\small jl4284@cornell.edu}
}

\twocolumn[{
\maketitle
\begin{center}
    \captionsetup{type=figure}
    \vspace{-0.6cm}
    \includegraphics[width=1\textwidth]{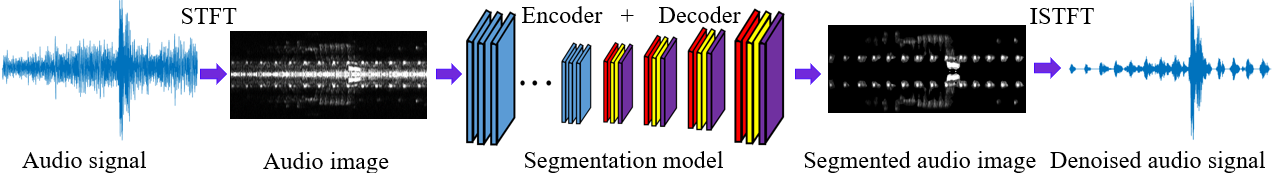}
    \captionof{figure}{The overall progress of our proposed deep visual audio denoising model (DVAD).} \label{fig:dvad}
\end{center}
\vspace{+0.1cm}
}]

\maketitle
\ifwacvfinal\thispagestyle{empty}\fi


\begin{abstract}
Audio denoising has been explored for decades using both traditional and deep learning-based methods. However, these methods are still limited to either manually added artificial noise or lower denoised audio quality. To overcome these challenges, we collect a large-scale natural noise bird sound dataset. We are the first to transfer the audio denoising problem into an image segmentation problem and propose a deep visual audio denoising (DVAD) model. With a total of 14,120 audio images, we develop an audio ImageMask tool and propose to use a few-shot generalization strategy to label these images. Extensive experimental results demonstrate that the proposed model achieves state-of-the-art performance. We also show that our method can be easily generalized to speech denoising, audio separation, audio enhancement, and noise estimation. 
\end{abstract}

\vspace{-0.3cm}
\section{Introduction}
With the development of technology, audio signals have been increasingly used as main sources of information transmission~\cite{li2018principal}, such as teleconferences~\cite{kong2021speech}, the speech-to-text function in social media~\cite{tamura2019novel}, the lung~\cite{pouyani2022lung} and heart~\cite{kui2021heart} sounds for disease diagnosis, instrument solo identification~\cite{gomez2018jazz}, and hearing aid~\cite{wang2021denoising,saleem2020deep,tan2018convolutional}, etc. Therefore, it is important to maintain the quality of signal transmission and retain as much useful information as possible. However, due to existing noises in the actual environment, the transmission of audio signals, including speech and other signals that we intend to collect, are inevitably affected, resulting in the poor quality and intelligibility of audio signals. Audio denoising can significantly increase audio quality and contribute to a better outcome of information transition. 

Audio denoising has been a popular research area in recent years and different methods have been applied to reduce noise and separate audio, including traditional statistics approaches~\cite{boll1979suppression,scalart1996speech,hansen2006speech,martin2005speech} and deep learning approaches~\cite{saleem2020deep,saleem2019deep,alamdari2021improving,raj2021multilayered,li2020speech}. While there are several difficulties encountered across these models. In this paper, we specifically use samples from the natural environment, which presents more challenges to the proposed research models. 

\hspace{-0.4cm}\textbf{Why natural audio denoising is difficult?} 

Firstly, the most common difficulty encountered is the limited sources for training. Deep learning-based models require both clean and noisy audio samples for training. However, in reality, audio signals come with noises that cannot be separated to produce desired training samples~\cite{kong2021speech}. Secondly, most noisy audio samples used for model training are artificially compiled, such as white gaussian noise (WGN)~\cite{srivastava2016new,xu2014regression}, which is composed differently from natural noise. In addition, we could still observe the clean signal patterns in the artificial noise audio, while it is difficult to observe the clean signal patterns in real noise audio as shown in Fig.~\ref{fig:dvad} (leftmost and rightmost signal). Therefore, the denoising performance of the training models might not perform as well in the real setting compared to experiments. 

These two challenges are commonly encountered in the audio denoising field, and we address them using a deep visual audio denoising model (DVAD). In this paper, we first collect audio samples that are directly acquired from the natural environment. The proposed model can process more complex and natural noises compared to previous models. We offer three principal contributions:
\begin{itemize}[noitemsep,topsep=0pt]
    \item We present a benchmark bird sounds denoising dataset with the goal of advancing the state-of-the-art in audio denoising under natural noise background.

    \item To the best of our knowledge, we are the first to transfer audio denoising into an image segmentation problem. By removing the noise area in the audio image, we can realize the purpose of audio denoising.
    
    \item We develop an audio ImageMask tool to label the collected dataset and apply a few-shot generalization strategy to accelerate the data label process. We also demonstrate that our model can be easily extended to speech denoising, audio separation, audio enhancement, and noise estimation.
    
\end{itemize}

\section{Related Work}
Audio denoising has been widely explored, and many methods have evolved from traditional methods of estimating the difference between noise and clean audio statistics~\cite{wang2021denoising}, to the adoption of deep learning methods~\cite{azarang2020review}.

Traditional methods for audio denoising can be dated back to the 1970s. Boll~\cite{boll1979suppression} proposed a noise suppression algorithm for spectral subtraction using the spectral noise bias calculated in a non-speech environment. Another statistical method proposed in~\cite{scalart1996speech} is a more comprehensive algorithm, combining the concept of A Priori Signal-to-Noise Ratio (SNR) with earlier typical audio enhancement schemes such as Wiener filtering~\cite{chen2006new,lim1979enhancement}, spectral subtraction, or Maximum Likelihood estimates. In the realm of the frequency-domain algorithm, minimum mean square error (MMSE) based approaches is a mainstream approach besides Wiener filtering. Hansen et al.~\cite{hansen2006speech} proposed an auditory masking threshold enhancement method by applying Generalized MMSE estimator in an auditory enhancement scheme. In~\cite{martin2005speech}, the MMSE estimator is used to enhance the performance of short-time spectral coefficients by estimating discrete Fourier transform (DFT) coefficients of both noisy and clean speech. One major problem is that the performance of traditional methods for noise separation and reduction will be degraded with the presence of natural noises, which are largely different from artificial noises applied in the experiments.

Wavelet transformation methods are developed to overcome the difficulty of studying signals with low SNR and are reported with better performance than filtering methods. Zhao et al.~\cite{zhao2015improved} used an improved threshold denoising method, overcame the discontinuity in hard-threshold denoising, and reduced the permanent bias in soft-threshold denoising. Srivastava et al.~\cite{srivastava2016new} developed a wavelet denoising approach based on wavelet shrinkage, allowing for the analysis of low SNR signals. Pouyani et al.~\cite{pouyani2022lung} proposed an adaptive method based on discrete wavelet transform and artificial neural network to filtrate lung sound signals in a noisy environment. Kui et al.~\cite{kui2021heart} also combined the wavelet algorithm with CNNs to classify the log mel-frequency spectral coefficients features from the heart sound signal with higher accuracy. These combined methods outperformed single wavelet transformation methods. 

Deep learning methods are later introduced to the audio denoising field, complementing the disadvantages of traditional methods and demonstrating a stronger ability to learn data and characteristics with a few samples~\cite{wang2021denoising}. The deep neural network (DNN)-based audio enhancement algorithms have shown great potential in their ability to capture data features with complicated nonlinear functions~\cite{li2020speech}. Xu et al.~\cite{xu2020listening} introduced a deep learning model for automatic speech denoising to detect silent intervals and better capture the noise pattern with the time-varying feature. Saleem et al.~\cite{saleem2020deep} used the deep learning-based approach for audio enhancement accompanying complex noises. An ideal binary mask (IBM) is used during the training and testing, and the trained DNNs are used to estimate IBM. Xu et al.~\cite{xu2014regression} proposed a DNN-based supervised method to enhance audio by finding a mapping function between noisy and clean audio samples. A large mixture of noisy dataset is used during the training and other techniques, including global variance equalization and the dropout and noise-aware training strategies. Saleem et al.~\cite{saleem2019deep} also developed a supervised DNN-based single channel audio enhancement algorithm and applied less aggressive Wiener filtering as an additional DNN layer. Vuong et al.~\cite{vuong2021modulation} described a modulation-domain loss function for a deep learning-based audio enhancement approach, applying additional Learnable spectro-temporal receptive fields to enhance the objective prediction of audio quality and intelligibility. 

Yet a problem in speech denoising application of DNN is that sometimes, it is difficult for models to track a target speaker in multiple training speakers, which means that the DNNs are not easy to handle long-term contexts~\cite{tan2018convolutional,li2020speech}. Therefore, deep learning approaches, such as convolutional neural network (CNN)-based and recurrent neural network (RNN)-based models, are explored. Alamdari et al.~\cite{alamdari2021improving} applied a fully convolutional neural network (FCN) for audio denoising with only noisy samples, and the study displayed the superiority of the new model compared to the traditional supervised approaches. Germain et al.~\cite{germain2018speech} trained an FCN using a deep feature loss, trained for acoustic environment detection and domestic audio tagging. The research showed that this new approach is particularly useful for audio with the most intrusive background noise. Kong et al.~\cite{kong2021speech} proposed an audio enhancement method with pre-trained audio neural networks using weakly labeled data with only audio tags of audio clips and applied a convolutional U-Net to predict the waveform of individual anchor segments selected by PANNs.
Raj et al.~\cite{raj2021multilayered} proposed a multilayered CNN-based auto-CODEC for audio signal denoising, using the mel-frequency cepstral coefficients, providing good encoding and high security. Abouzid et al.~\cite{abouzid2019signal} combined the convolutional and denoising autoencoders into convolutional denoising autoencoders for the suppression of noise and compression of audio data.  

On top of single-type deep learning methods, Tan et al.~\cite{tan2018convolutional} proposed a recurrent convolutional network by incorporating a convolutional encoder-decoder and long short-term memory (LSTM) into the  convolutional recurrent neural network (CRN) architecture to address real-time audio enhancement. This method outperformed an existing LSTM-based model with fewer trainable parameters. Gao et al.~\cite{gao2016snr}, and ~\cite{gao2018densely} respectively applied a progressive learning framework for DNN-based and LSTM-based audio enhancement to improve model performance and reduce complexity. Li et al.~\cite{li2020speech}  combined the progressive learning framework with a causal CRN to further reduce the trainable parameters and improve audio quality and intelligibility. This proposed method produced a close performance to the CRN. 

Many deep learning approaches are implemented in the time-frequency domain, using short-time Fourier transform (STFT) and inverse short-time Fourier transform (ISTFT)~\cite{wang2021tstnn}. Some methods address audio enhancement via time-domain algorithms, viewing audio enhancement as a filtering problem ~\cite{yu2019deep}. Yu et al.~\cite{yu2019deep} proposed a DNN-based Kalman filter algorithm for audio enhancement. The DNN is used for estimating the linear prediction coefficients in the KF. Sonning et al.~\cite{sonning2020performance} investigated the performance of a time-domain network for speech denoising, addressing the original inability of STFT/ISTFT-based time-frequency approaches to capture short-time changes and discovered its usefulness in a real-time setting. Wang et al.~\cite{wang2021tstnn} proposed a two-stage transformer neural network  for end-to-end audio denoising in the time domain, including an encoder, a two-stage transformer module , a masking module and a decoder. Their model outperformed many time- or frequency-domain models with less complexity.

\section{Methods}
\subsection{Problem}
Given a noisy audio signal $\{x_t\}_{t=1}^{T}$, we aim to extract the clean audio $\{y_t\}_{t=1}^{T}$ by learning a mapping $\mathcal{M}$.  The goal of audio denoising is to minimize the approximation error between the denoised audio $\{\mathcal{M}(x_t)\}_{t=1}^{T}$ and clean audio $\{y_t\}_{t=1}^{T}$. In our DVAD model, we convert audio denosing to an image segmentation problem. Given the audio images $\mathcal{I} = \{I^i\}_{i=1}^{n}$ based on audio signals $X = \{x^i\}_{i=1}^{n}$ and its ground truth labeled masks $M =  \{m^i\}_{i=1}^{n}$, we propose to minimize the error between prediction of any image segmentation model $F(I)$ and $M$.

\subsection{Motivation}
Although some existing deep audio denoising models utilized magnitude images of audio signals, they only filtered out some regions of the image to realize the purpose of denoising. The details of these images are less explored. Our DVAD model delves into the audio image to find different patterns between noise and clean signal areas. As shown in Fig.~\ref{fig:b}, we can find that there are significant differences between the noise and clean signal areas. Therefore, we can achieve the purpose of audio denoising if we can segment the clean signal areas. We further treat the audio denoising as an image segmentation problem.

\subsection{Preliminary}
\subsubsection{Short-Time Fourier Transform (STFT)}
STFT is used to analyze how the frequency content of a nonstationary signal changes over time.
\begin{equation}\label{eq:STFT}
    STFT_x(t,f) = \int_{-\infty}^{\infty} x(t) \omega(t-\tau) e^{-j2\pi f t} dt
\end{equation}
where $STFT_x(t,f)$ is the coefficient of STFT. STFT is a function of time ($t$) and frequency ($f$), and it shows how frequency $f$ of the signal $x(t)$ changes with time $t$. $\omega$ is a window function, $\tau$ is a short time and $j$ is the square root of $-1$. In our model, we aim to convert the signal to frequency domain and get the raw images for each bird sound. 

\subsubsection{Inverse Short-Time Fourier Transform (ISTFT)}
The STFT is invertible, $i.e.$, the original signal can be reconstructed from the transform by the inverse STFT. It is defined as:
\begin{equation}\label{eq:ISTFT}
    \Tilde{x_t} =  \int_{-\infty}^{\infty}\int_{-\infty}^{\infty} STFT_x(t',f') \omega(t-t') e^{-j2\pi f' t'} dt'\ df'.
\end{equation}
In our model, we aim to reconstruct the bird sound based on the segmented bird sound image.

\begin{figure}[t]
\centering
\begin{subfigure}{0.35\textwidth}
\includegraphics[width=\linewidth]{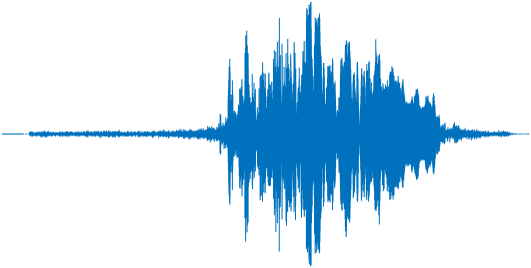}
\caption{Audio signal}
\end{subfigure}
\begin{subfigure}{0.12\textwidth}
\includegraphics[width=\linewidth]{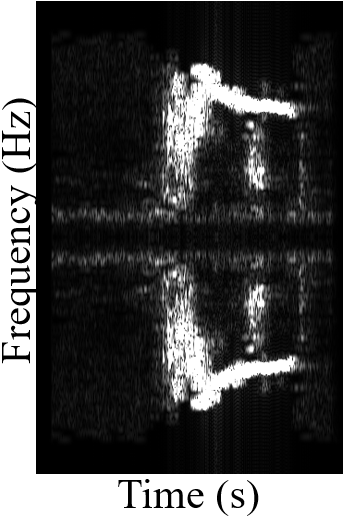}
\caption{Audio image}\label{fig:b}
\end{subfigure}
\vspace{-0.3cm}
\caption{The conversion from audio signal (a)  to audio image (b) by taking the absolute value of STFT. (b) is a symmetrical image, and most noise areas are concentrated in the center of an image (the light white horizontal lines). The informative and clean signal (bird sound) majorly lies in the bright white patterns. Please refer to supplementary material for more noise signal areas. } \label{fig:audio_images}
\vspace{-0.6cm}
\end{figure}

\subsection{Methodology}\label{sec:method}
\begin{figure}[h]
\centering
\begin{subfigure}{0.15\textwidth}
\includegraphics[width=\linewidth]{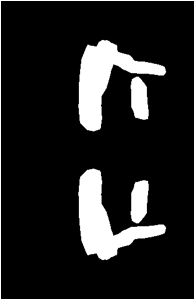}
\caption{Mask image}
\end{subfigure}
\hspace{+1cm}
\begin{subfigure}{0.15\textwidth}
\includegraphics[width=\linewidth]{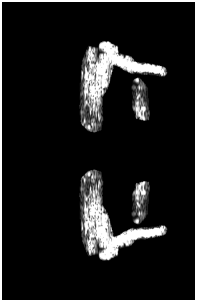}
\caption{Segmented  image}\label{fig:d_b}
\end{subfigure}
\vspace{-0.3cm}
\caption{The mask (a)  and segmented audio image (b) for the signal in Fig.~\ref{fig:audio_images}. (a) is the mask of clean signal areas in Fig.~\ref{fig:b} and (b) is the segmented audio image by removing noise areas. } \label{fig:mask_images}
\vspace{-0.3cm}
\end{figure}

To form the audio denoising problem as an image segmentation problem, we first need to represent the audio in image format. After performing STFT using Eq.~\eqref{eq:STFT}, let $S = STFT_x(t,f)$,  we can define the audio image ($I$) in the following equation,
\begin{equation}\label{eq:I}
    I = abs(S),
\end{equation}
where $abs$ takes the absolute value from the complex frequency domain $S$. As shown in Fig.~\ref{fig:audio_images}, we convert a one-second bird sound audio to its audio image. We observe that the patterns of noise areas and clean sound areas are distinguishable. If we segment the clean sound areas in the audio image, then we could remove the noise from the frequency domain $S$. As shown in Fig.~\ref{fig:d_b}, we can apply ISTFT in Eq.~\eqref{eq:ISTFT} after getting the denoised audio image to reconstruct the denoised audio signal. Therefore, we can convert the audio denoising problem into an image segmentation problem.  

To realize the image segmentation task, we need to train a segmentation model $F$ to segment the clean audio signal area. In our DVAD model, we train the segmentation model using dice loss as follows.
\begin{equation}\label{eq:dice}
    Dice \ loss = 1- 2 \times \frac{m \cap \Tilde{m}}{m + \Tilde{m}},
\end{equation}
where $m$ is the ground truth mask and $\Tilde{m} = F(I)$ is the predicted mask of the segmentation model given the input image $I$. In the mask, we denote the clean audio areas using 1 and represent the noise areas using 0. After training $F$, we can predict the segmented mask of any audio image. Next, we aim to reconstruct the denoised audio.  

\begin{figure}[h]
\centering
\begin{subfigure}{0.2\textwidth}
\includegraphics[width=\linewidth]{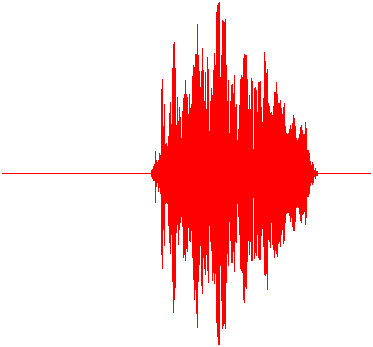}
\caption{Denosied audio}\label{fig:4a}
\end{subfigure}
\hspace{+1cm}
\begin{subfigure}{0.2\textwidth}
\includegraphics[width=\linewidth]{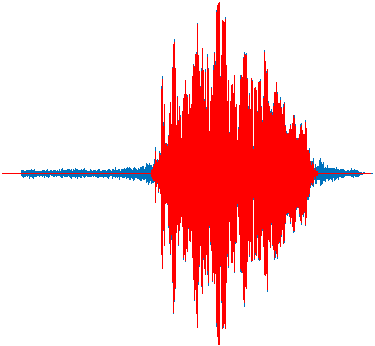}
\caption{Overlapping signals}\label{fig:4b}
\end{subfigure}
\vspace{-0.3cm}
\caption{The denoised audio using our DVAD model (a) and overlapping of original signal (blue color) and denoised audio signal (red color). } \label{fig:overlap}
\vspace{-0.3cm}
\end{figure}

In Eq.~\eqref{eq:ISTFT}, we can recover the original $x(t)$ given the key input frequency domain $S$. To remove the noise audio, we need to filter out noise areas in $S$ given the predicted mask from the segmentation model. We define the new frequent domain $S'$ in the following equation: 
\begin{equation}\label{eq:s'}
    S' = S, \ \ \  \text{and} \ \ \ S'[\Tilde{m} < 1] = 0,
\end{equation}
where $S'[\Tilde{m} < 1] = 0$ aims to replace all noise area with 0 to realize the purpose of removing noise areas. Then, we can apply ISTFT to reconstruct the denoised audio as follows.
\begin{equation}\label{eq:d_signal}
        \Tilde{x_t} =  \int_{-\infty}^{\infty}\int_{-\infty}^{\infty} S' \omega(t-t') e^{-j2\pi f' t'} dt'\ df'.
\end{equation}
As shown in Fig.~\ref{fig:4a}, we can get the denoised audio after removing the noise audio using Eq.~\eqref{eq:s'}. We also show the overlapping of original signal with denoised signal in Fig.~\ref{fig:4b}. The rest part of blue signal (noise areas) is removed from the red denoised signal.  

\subsection{DVAD overall algorithm}
Considering all steps in Sec.~\ref{sec:method}, the scheme of our proposed DVAD model is shown in Fig.~\ref{fig:dvad} and the overall algorithm is presented in Alg.~\ref{alg:DVAD}.
\begin{algorithm}[h]
   \caption{Deep Visual Audio Denoising (DVAD). $B(\cdot)$ denotes the mini-batch training sets, $I$ is the number of iterations.}
   \label{alg:DVAD}
\begin{algorithmic}[1]
   \STATE {\bfseries Input:} Audio signals  $X = \{x^i\}_{i=1}^{n}$ and labeled mask images $M =  \{m^i\}_{i=1}^{n}$, where $n$ is the total number of audios.
   \STATE {\bfseries Output:} Denoised audio signals
   \STATE Generate audio images $\mathcal{I} = \{I^i\}_{i=1}^{n}$ using Eq.~\eqref{eq:I}
   \FOR{$iter =1$ {\bfseries to} $I$}
   \STATE Derive $B(\mathcal{I})$ and $ B(M)$ sampled from $\mathcal{I}$ and $M$
   \STATE Optimize any segmentation model $F$ using Eq.~\eqref{eq:dice}
   \ENDFOR
   \STATE Get the clean frequency domain using Eq.~\eqref{eq:s'}
   \STATE Output the denoised audio signals using Eq.~\eqref{eq:d_signal}
\end{algorithmic}
\end{algorithm}

\begin{figure*}[t]
    \centering
    \includegraphics[width=1\textwidth]{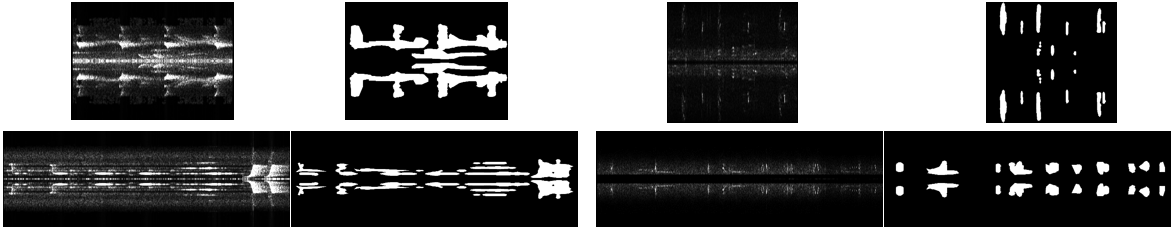}
    \caption{Four sample images and masks. In each sample, the left is the audio image, and the right is its labeled mask. The width of each sample depends on the length of the audio. Longer audio will produce a wider audio image. }
    \label{fig:samples}
    \vspace{-0.3cm}
\end{figure*}

\vspace{-0.6cm}
\section{Datasets}
\subsection{Data Collection}
Our data is collected from the xeno-canto website, which is a public website to share bird sounds from around the world \footnote{\url{https://xeno-canto.org/explore}}. We first collect 15,300 bird sounds from one second to fifteen seconds. Unlike many audio denoising datasets, which have manually added artificial noise, our collected bird sounds contain natural noises, including wind, waterfall, rain, etc. Then, we apply STFT in Eq.~\eqref{eq:STFT} to convert bird sounds to the frequency domain and get the audio images using Eq.~\eqref{eq:I}. Some audios have two soundtracks (left and right soundtracks). Hence we get more images than collected audios. Since we have converted audio denoising to an image segmentation problem, we need to provide masks for audio images to optimize any image segmentation models. Therefore, our next task is to label these audio images.

\subsection{Data labeling}\label{sec:label}
Image mask labeling is time-consuming, tedious, and expensive. However, to train any machine learning algorithms, we have to provide enough labeled datasets to achieve a good performance. Given that there is no specific software for audio image labeling, we also developed an audio ImageMask tool. The ImageMask software has three key functions.
(1). It can open an audio image and label it to create a mask and show the overlapping between the raw audio image and labeled mask.
(2). We can save created masks and denoised audios. The software also supports human verification. All accepted denoised audio will be saved in an `Accepted' folder. The folder contains another four sub-folders: original audio, denoised audio, audio images, and audio masks.
(3). We can also compare the ground truth masks with predicted masks from any segmentation models to validate the performance of models. More details of our developed audio ImageMask tool can be found in the supplementary material.

\hspace{-0.4cm}\textbf{Few-shot generalization} 

Although we developed a specific audio denoising software, it takes around 5 minutes to label one audio image. We have more than 15,000 audio images, and it is still time-consuming to label all images. To accelerate the labeling process, we propose to utilize the few-shot generation to first predict coarse masks for audio images. Then we can verify and update these coarse masks to get better masks.  

Few-shot learning aims to learn a robust model based on a few labeled samples, then improve the performance of new datasets. To ease the process of image labeling from scratch, we first manually labeled 100 audio images as training and 40 images as test. We select DeepLabV3~\cite{chen2018encoder} as the segmentation model and train the DeepLabV3 model using these 140 labeled images to get a basic model $F$. We could then predict the coarse mask via $F(I)$. Given any unlabeled audio image $I^i$, we can get all predicted coarse masks as $\{F(I^i)\}_{i=1}^{n}$. Finally, these coarse masks can be further modified using our developed audio ImageMask tool. After using the proposed few-shot generalization strategy, the whole dataset is labeled by four experts in one month.

\subsection{BirdSoundsDenoising dataset}
After finishing the data labeling process using Sec.~\ref{sec:label}, we can save all accepted labeled audio to create the BirdSoundsDenoising dataset (note that some low-quality audio will be removed during the labeling process). The BirdSoundsDenoising dataset contains 14,120 audio images and has three folders: training, validation, and test. In each folder, another four sub-folders are included: raw audios, denoised audios, images, and masks. Tab.~\ref{tab:data} shows the statistics of each folder. As shown in Fig.~\ref{fig:samples}, we list four audio images and their labeled masks. These audio images are varied, and clean signal areas are also different. Longer audio can produce a longer image. In the experiments, we will fine-tune models using training and validation datasets and report results on the test dataset.
\begin{table}[h]
\footnotesize
\begin{center}
\caption{Statistics on BirdSoundsDenoising dataset}
\vspace{-0.3cm}
\setlength{\tabcolsep}{+4mm}{
\begin{tabular}{rccccccccccccc}
\hline \label{tab:data}
Datasets &  Training & Validation & Test \\
\hline
Number of samples &  10,000 & 1,400 & 2,720 \\
 \hline
\end{tabular}}
\end{center}
\vspace{-0.6cm}
\end{table}

\subsection{Dataset creation details}\label{sec:data_label}
In STFT, converting audio signals to audio images, we first use 128-point Hamming as the window function, the number of overlapped samples $= 64$, the number of DFT points $=1024$, to change bird sounds to frequency domain $S$, and save all bird audio images using Eq.~\eqref{eq:I}. In few-shot generalization, we utilize DeepLabV3 as a basic segmentation model to generate coarse audio image masks. During the training of 140 images, we set batch size = 16, training iteration $I = 100$, learning rate = 0.0001 with an Adam optimizer on a RTX A6000 GPU. The input image size of the DeepLabV3 model is $[512 \times 512 \times 3]$. We excluded audio images with no bird sound or extreme noisy background.

\section{Experiments}
To evaluate the performance of our proposed DVAD model, we test it on our created BirdSoundsDenoising dataset. We first train six different state-of-the-art segmentation models to demonstrate the effects of different segmented masks on audio denoising performance. These six selected segmentation models have an encoder-decoder architecture. The encoder aims to extract important features from images (e.g., edge), and the decoder learns how to map these low-resolution features to the prediction at the pixel level.
\begin{enumerate}
   \item SegNet~\cite{badrinarayanan2017segnet}: the encoder network utilizes the layers from VGG16, and the decoder network is followed by a pixel-wise classification layer.  The decoder of SegNet can upsample its lower-resolution input feature map(s) using pooling indices computed in the maxpooling step of the corresponding encoder to perform non-linear up-sampling.
    
    \item U-Net~\cite{ronneberger2015u}: it has a structure called dilated convolutions and removes the pooling layer structure. The U-Net architecture comprises a contracting path to capture context, and a symmetric expanding path that enables precise localization.
    
    \item DeepLabV3~\cite{chen2018encoder}: it uses dilated convolutions and a fully connected conditional random field to implement the atrous spatial pyramid pooling (ASPP), which is an atrous version of SPP and can account for different object scales and improve the accuracy.
    
    \item U$^2$-Net~\cite{qin2020u2}: it is able to capture more contextual information from different scales with the mixture of receptive fields of different sizes in ReSidual U-blocks.

    \item Segmenter~\cite{strudel2021segmenter}: it utilizes the Vision Transformer (ViT) as the encoder to encode all image patches. A point-wise linear decoder is applied to patch encodings. It also includes a decoder with a mask transformer to further improves the performance.

    \item MTU-Net~\cite{wang2022mixed}: it proposes a novel transformer module named Mixed Transformer Module (MTM). It calculates self affinities efficiently by a Local-Global Gaussian-Weighted Self-Attention (LGG-SA). It also mines inter-connections between data samples using External Attention (EA).
\end{enumerate}

\subsection{Implementation details}
The training settings of six different segmentation methods are the same as Sec.~\ref{sec:data_label} except that we use batch size of 12 for MTUNet and  32 for Segmenter\footnote{BirdSoundsDenoising dataset and code are available at \url{https://github.com/YoushanZhang/BirdSoundsDenoising}}. To show the superiority of the DVAD model, we also compare the proposed model with three audio denoising methods ~\cite{park2017fully,kashyap2021speech,moliner2022two}.

\begin{figure*}[t]
    \centering
    \includegraphics[width=0.9\textwidth]{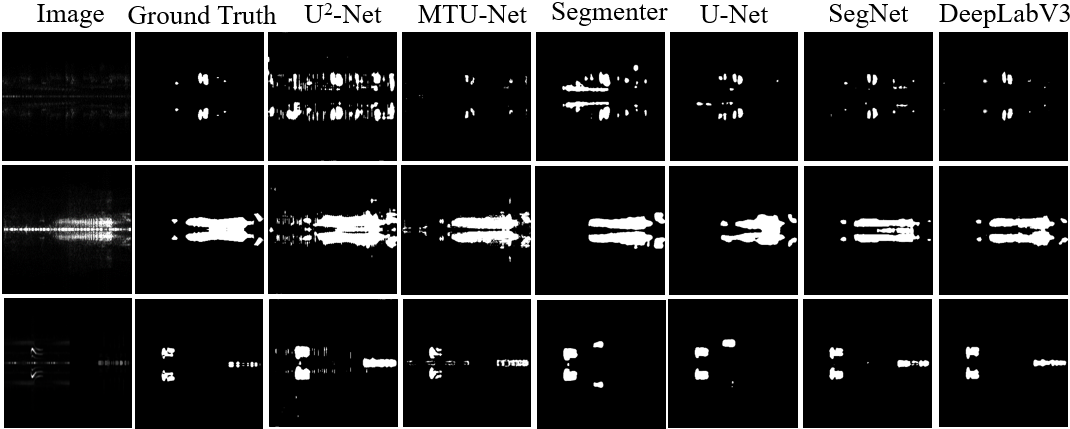}
    \caption{Segmentation results comparisons. Leftmost column is the original audio image. Ground truth is the labeled mask. }
    \label{fig:seg_re}
    \vspace{-0.3cm}
\end{figure*}
\begin{figure*}[t]
    \centering
    \includegraphics[width=0.9\textwidth]{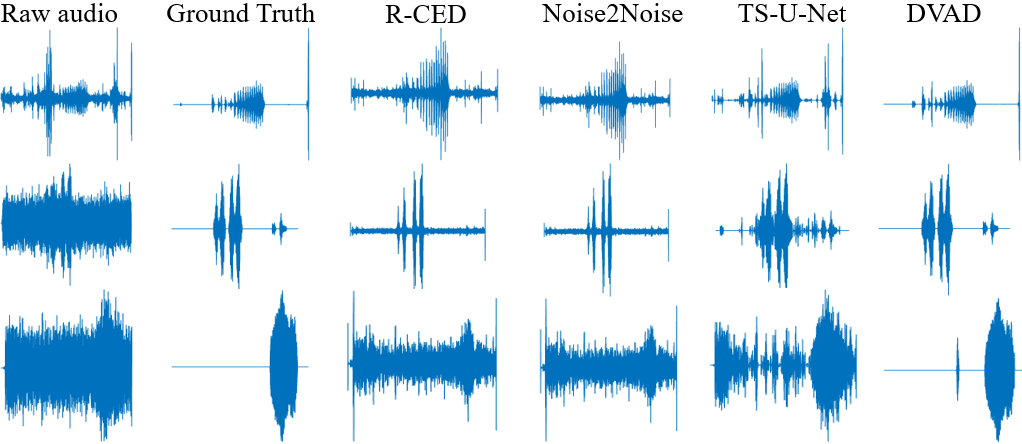}
    \caption{Denoising results comparisons. Raw audio is the original noise audio. Ground truth is the labeled mask. }
    \label{fig:sig_re}
    \vspace{-0.6cm}
\end{figure*}

\subsection{Results}
\subsubsection{Evaluation metrics}
We use three metrics ($F1, IoU$ and $Dice$) to evaluate the performance of image segmentation as follows. 
\begin{equation}
\begin{aligned}
    & F1 = \frac{TP}{TP + \frac{1}{2} (FP+FN)}, \\ & IoU =  \frac{m \cap \Tilde{m}}{m + \Tilde{m}} \ \ \ \ Dice =  2 \times \frac{m \cap \Tilde{m}}{m + \Tilde{m}},    
\end{aligned}
\end{equation}
where $TP$ is the number of true positives,  $FP$ is the number of false positives, and $FN$ is the	number of false negatives. For audio denoising, we use signal-to-distortion ratio (SDR) to evaluate our DVAD model. The higher these four metrics, the better the segmentation model is.
\begin{equation}
    SDR = 10\  \text{log}_{10}\frac{||m||^2}{||\Tilde{m} - m||^2}
\end{equation}

\subsection{Performance comparisons}
We first show the comparisons of six different segmentation models in Fig.~\ref{fig:seg_re}. The segmented masks of the DeepLabV3 model are better than the other five models. Similarly, we could observe that DeepLabV3 has the highest $F1, IoU$, and $Dice$ scores in Tab.~\ref{tab:md}. Therefore, we can infer that the DeepLabV3 model is the best segmentation model for our BirdSoundsDenoising among all six segmentation models. In addition, we also reported the mean SDR of all bird sounds in both validation and test datasets. As shown in Tab.~\ref{tab:md}, the SDR score of our DVAD model with DeepLabV3 as the segmentation model achieves the highest value. Notably, the performance level of three audio denoising methods (R-CED, Noise2Noise, and TS-U-Net) is relatively lower than all other segmentation models. The comparisons of raw bird audio, true labeled denoised audio, and denoised audio from other models are shown in Fig.~\ref{fig:sig_re}. The denoised signal of DVAD (with DeepLabV3) is also closer to the labeled denoised signal. Therefore, our DVAD architecture is effective in improving the audio denoising performance.

\section{Discussion}
We compared six different state-of-the-art segmentation models and three deep audio denoising methods. One obvious strength of our model is its better performance than other methods. Especially the different variants of our DVAD model are significantly better than the three audio denoising methods in terms of SDR score. The compelling advantage of the DVAD model lies in the image segmentation section. We can maintain the crucial clean signal via a segmented mask, as shown in Fig.~\ref{fig:seg_re}. Given a clean area in the segmented masks, the clean signal will be preserved during the ISTFT process. Therefore, converting audio denoising into an image segmentation problem can be a new stream to further improve the performance of audio denoising. 

\begin{table}[t]
\small
\begin{center}
\captionsetup{font=small}
\caption{Results comparisons of different methods ($F1, IoU$, and $Dice$ scores are multiplied by 100. ``$-$" means not applicable. }
\vspace{-0.3cm}
\setlength{\tabcolsep}{+0.3mm}{
\begin{tabular}{lllll|lllllllll}
\hline \label{tab:md}
 \multirow{2}{*}{Networks}
 &  \multicolumn{4}{c}{Validation} & \multicolumn{4}{c}{Test} \\
 \cmidrule{2-9}
& $F1$ & $IoU$  & $Dice$ & $SDR$ & $F1$ & $IoU$  & $Dice$ & $SDR$ \\
\hline
U$^2$-Net~\cite{qin2020u2}  &60.8 &45.2 &60.6 &7.85 & 60.2  &44.8 &59.9 & 7.70\\
MTU-NeT~\cite{wang2022mixed}  &69.1 &56.5 &69.0  &8.17 & 68.3  &55.7 & 68.3 &7.96  \\
Segmenter~\cite{strudel2021segmenter} & 72.6  & 59.6 & 72.5 & 9.24 & 70.8 & 57.7 & 70.7 & 8.52   \\
U-Net~\cite{ronneberger2015u}  &75.7 &64.3 &75.7 & 9.44 &74.4 &62.9 &74.4 & 8.92    \\
SegNet~\cite{badrinarayanan2017segnet}  &77.5 &66.9 &77.5 & 9.55&76.1 &65.3 &76.2 & 9.43 \\
DeepLabV3~\cite{chen2018encoder}  & \textbf{82.6}  & \textbf{73.5} & \textbf{82.6} &  \textbf{10.33}  & \textbf{81.6} & \textbf{72.3} & \textbf{81.6} & \textbf{9.96} \\
\hline  
\hline
R-CED~\cite{park2017fully} & $-$ & $-$ & $-$ &2.38     &$-$ &$-$&$-$ & 1.93  \\
Noise2Noise~\cite{kashyap2021speech}  & $-$ & $-$ & $-$ & 2.40&$-$ &$-$&$-$ &1.96\\
TS-U-Net~\cite{moliner2022two}  & $-$ & $-$ & $-$ & 2.48&$-$ &$-$&$-$ &1.98\\
\hline
\end{tabular}}
\end{center}
\vspace{-.9cm}
\end{table}

\section{Conclusion}
In this paper, we are the first to convert audio denoising into an image segmentation problem. We then propose a deep visual audio denoising (DVAD) network to remove the noise from a larger-scale BirdSoundsDenoising dataset. In addition, we design an audio ImageMask tool and propose to use few-shot generation to label all datasets. Extensive experimental results demonstrate that the proposed DVAD model outperforms many state-of-the-art methods. As for future work, a novel segmentation model could be developed to further improve the audio denoising performance.

\section{Broad impact}
The application of our proposed DVAD model is not limited to bird sound denoising. It can be easily extended to the following tasks for real-life applications. 

\subsection{Adaptation to speech denoising}
\begin{figure}[h]
    \centering
    \includegraphics[width=0.47\textwidth]{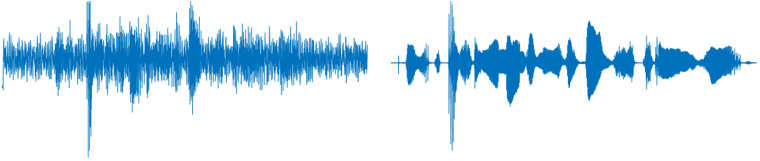}
     \caption{Example of speech denoising. Left is the original speech with noise and the right is the denoised speech audio. }
    \label{fig:speech}
    \vspace{-0.3cm}
\end{figure}

Speech denoising is increasingly important as speech has become a predominant medium of daily communication and the main aspect of technology advancement. We use bird sounds as training samples for our model, but our model can also be applied to human speech denoising or even other non-audio signals. We applied the pre-trained DeepLabV3 model to segment the speech audio image, then converted the segmented image to get the denoised speech audio. As shown in Fig.~\ref{fig:speech}, the noise in human speech audio can be significantly reduced. Therefore, our DVAD model demonstrates high-quality performance in human speech audio denoising.

\subsection{Audio separation}
\begin{figure}[h]
    \centering
    \includegraphics[width=0.45\textwidth]{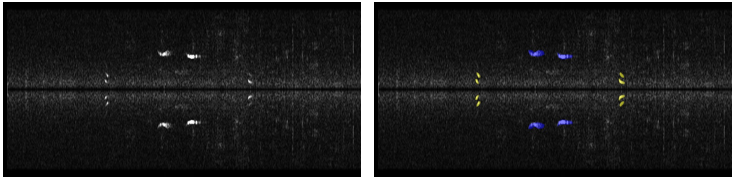}
    \caption{Example of audio separation. Left is the audio image and right is the overlay of audio image with two different segmented bird masks. Yellow color is one bird, while blue color is another bird.}
    \label{fig:audio_seperation}
\end{figure}

Audio separation tasks are important in many scenarios, such as high-quality conference video production, surveillance system use, audio identification, etc. In the birds sound denoising task, we treat it as a binary image segmentation problem. For audio separation, we could treat it as a multi-class image segmentation problem. As shown in Fig.~\ref{fig:audio_seperation}, we can separate two different bird sounds. We also added separated bird sounds in the supplementary material. This aspect of our model is significant when trying to detect or identify desired audio signal in a noisy mix. Hence, our DVAD model can be easily applied to the audio separation problem.   

\subsection{Audio enhancement}
\begin{figure}[h]
    \centering
    \includegraphics[width=0.47\textwidth]{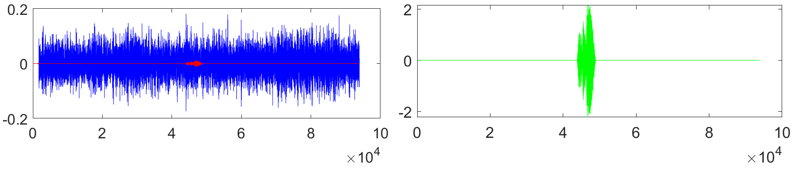}
     \caption{Example of audio enhancement. The blue line is the original noise signal, and the red line is the denoised signal. On the right, it is the enhanced signal of 200 times of signal in the red line on the left. x and y axes represent the length and magnitude of the signal, respectively.}
    \label{fig:audio_enhancement}
    \vspace{-0.3cm}
\end{figure}

Audio enhancement has always been a challenging task since the noise signal would also be enhanced if we did not properly remove noise signals. This challenge presents in many models designed for speech enhancement. In our DVAD model, we only preserve the clean signal mask. Hence, higher quality audio enhancement can be achieved. Given the denoised audio $\hat{x_t}$, we can enhance the audio by enlarging the denoised signal with $l\hat{x_t}$, where $l$ is the number of times to enlarge the signal. As shown in Fig.~\ref{fig:audio_enhancement}, we can enhance the pure bird signal by $l=200$ times. An enhanced audio example can also be found in the supplementary material. Audio enhancement has a much wider application, such as hearing aid, recording production, long-distance signal transmission like cellular communication, etc.

\begin{figure}[h]
    \centering
    \includegraphics[width=0.47\textwidth]{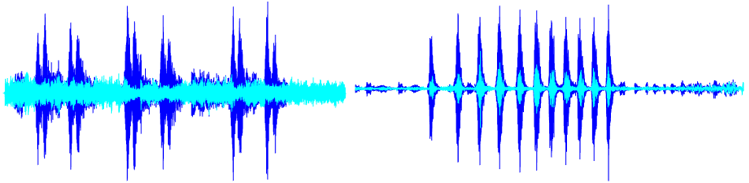}
     \caption{Two examples of noise estimation. The blue color is the noise audio, and the cyan color is the estimated noise.}
    \label{fig:audio_noise}
    \vspace{-0.3cm}
\end{figure}
\subsection{Noise estimation and audio identification}
In denoised audio, we could still occasionally hear the noise since the noise signal is difficult to be completely removed. Noise estimation will be useful if we can learn the patterns of noise, then we can further remove noise from the denoised audio. As shown in Fig.~\ref{fig:audio_noise}, we can estimate the noise signal by using original noise audio to subtract the clean denoised signal. Sampling these extracted noises can be used to further improve the quality of denoised signals. Noise estimation is particularly useful for model training, it could also be extended to audio identification. Noise in one scenario could become intended signals in others~\cite{xu2020listening}. As each audio signal has its own pattern, learning the pattern of different audio signals can be useful to match intended signals with those in training sets. This application is important in medical applications, e.g., identifying disease-likely sound patterns from regular sound patterns.


{\small
\bibliographystyle{ieee_fullname}
\bibliography{egbib}
}

\end{document}